\documentclass[12pt]{article}
\usepackage{epsfig,amssymb,amsmath,psfrag,hyperref}
\usepackage[utf8]{inputenc}
\usepackage[english]{babel}
 
\usepackage{color}

\hypersetup{
  linktocpage,
  colorlinks  = true, 
  urlcolor    = blue, 
  linkcolor   = blue, 
  citecolor   = red 
}

\def \be  {\begin{equation}}
\def \ee  {\end{equation}}
\def \ba  {\begin{eqnarray}}
\def \ea  {\end{eqnarray}}

\def \cO{\mathcal{O}}
\newcommand \widebar [1] {\overline{#1}}
\def\x{x}
\def\xb{\bar{x}}
\def\dbar#1{\widebar{D}_{#1}}
\def\me{\mathcal{M}}

\def\l{\lambda}
\def\ll{\lambda^{-\frac{3}{2}}}
\def\lll{\lambda^{-\frac{5}{2}}}

\def\fourtwo{\langle\cO_2\cO_2\cO_2\cO_2\rangle}
\def\sg{S}
\def\namb{N_{\text{amb}}}

\begin{document}
\thispagestyle{empty}

\null\vskip-12pt \hfill  \\
\null\vskip-12pt \hfill   \\

\vskip2.2truecm
\begin{center}
\vskip 0.2truecm {\Large\bf
{\Large One-loop string corrections to AdS amplitudes from CFT}
}\\
\vskip 1truecm
{\bf J.~M. Drummond and H.~Paul\\
}

\vskip 0.4truecm
 
{\it
 School of Physics and Astronomy, University of Southampton,\\
 Highfield, Southampton SO17 1BJ\\
\vskip .2truecm                        }
\end{center}

\vskip 1truecm 
\centerline{\bf Abstract}\normalsize
We consider $\alpha'$ corrections to the one-loop four-point correlator of the stress-tensor multiplets in $\mathcal{N}=4$ super Yang-Mills at order $1/N^4$. Holographically, this is dual to string corrections of the one-loop supergravity amplitude on AdS$_5\times$S$^5$.
While this correlator has been considered in Mellin space before, we derive the corresponding position space results, gaining new insights into the analytic structure of AdS loop amplitudes. Most notably, the presence of a transcendental weight three function involving new singularities is required, which has not appeared in the context of AdS amplitudes before. We thereby confirm the structure of string corrected one-loop Mellin amplitudes, and also provide new explicit results at orders in $\alpha'$ not considered before.

\medskip
\noindent                                   
\newpage
\setcounter{page}{1}\setcounter{footnote}{0}
\tableofcontents
\section{Introduction}\setcounter{equation}{0}
There has been much recent progress on using the AdS/CFT correspondence to understand the nature of AdS bulk gravitational dynamics by imposing consistency of the boundary conformal field theory. A particular example which admits a simple enough structure to allow for explicit results is the archetypal example of the AdS/CFT correspondence with boundary theory given by $\mathcal{N}=4$ super Yang-Mills theory in the large $N$ and strong 't~Hooft coupling limits, and the bulk theory given by IIB superstring theory on AdS${}_5 \times S^5$.

The natural objects to consider in the first instance are the four-point functions of half-BPS operators. In the supergravity regime, the large $N$ expansion corresponds to the loop expansion in the bulk theory. The leading order  is given by disconnected contributions, $1/N^2$ corrections correspond to tree-level bulk interactions and $1/N^4$ corrections to one-loop bulk contributions etc. At each order in $1/N^2$ the amplitude is a function of the 't~Hooft coupling $\lambda = g^2 N$ and we may further expand the loop amplitudes for large $\lambda$. The leading terms may be identified with supergravity contributions and subleading terms with string corrections.

This subject has been well studied since the very early days of the AdS/CFT correspondence. For example some explicit tree-level results were obtained from considering bulk Witten diagrams \cite{Freedman:1998tz,DHoker:1999kzh,DHoker:1999mqo} or related methods \cite{Arutyunov:2000py,Uruchurtu:2008kp,Uruchurtu:2011wh}. The AdS/CFT correspondence also sparked much activity in investigating the structure of the $\mathcal{N}=4$ superconformal theory \cite{Arutyunov:2000ku,Eden:2000bk,Dolan:2001tt,Nirschl:2004pa,Dolan:2004iy}. Among the more recent results on the subject is a general Mellin space formula for tree-level supergravity correlators of arbitrary external charges~\cite{Rastelli:2016nze,Rastelli:2017udc}. This was later confirmed by explicit computations in a large number of cases~\cite{Arutyunov:2017dti,Arutyunov:2018tvn}.

From tree-level data, one can analyse the spectrum of exchanged operators, which at leading order is a degenerate set of double-trace operators. After solving the mixing problem of supergravity anomalous dimensions~\cite{Aprile:2017xsp,Aprile:2018efk}, a surprisingly simple pattern emerges, revealing a partial residual degeneracy. This residual degeneracy in the supergravity spectrum is explained by the recent discovery of a hidden ten-dimensional conformal symmetry, which enables one to generate tree-level correlators of arbitrary charge half-BPS single-particle operators from a single generating functional~\cite{Caron-Huot:2018kta}. A similar structure seems also to be present for holographic tree-level correlators in AdS$_3$~\cite{Rastelli:2019gtj}.

Many new results concern further $\alpha'\sim\l^{-\frac{1}{2}}$ corrections to the supergravity results. Using various methods such as comparison to the flat-space limit~\cite{Goncalves:2014ffa,Alday:2018pdi} or supersymmetric localisation~\cite{Binder:2019jwn}, the family of $\langle\cO_2\cO_2\cO_p\cO_p\rangle$ correlators has been addressed up to order $\lll$. Interestingly, some of the ten-dimensional structure observed in~\cite{Caron-Huot:2018kta} seems to carry over to string corrected tree-level correlators. In particular, the flat space limit determines correlators with arbitrary external charges at order $\ll$, and simple patterns arise in the string corrected anomalous dimensions~\cite{Drummond:2019odu}.

In addition to the recent progress and emergence of beautiful structures at tree level, there have also been advances in understanding contributions to one-loop amplitudes in AdS (e.g. in ~\cite{Aharony:2016dwx,Giombi:2017hpr,Yuan:2017vgp,Yuan:2018qva,Carmi:2019ocp}). Recently a number of papers have addressed the structure of one-loop IIB supergravity amplitudes in AdS${}_5\times S^5$ ~\cite{Alday:2017xua,Aprile:2017bgs,Aprile:2017qoy,Aprile:2019rep,Alday:2019nin}. Progress has been made on two fronts, with e.g. \cite{Aprile:2017bgs,Aprile:2017qoy,Aprile:2019rep} focussing on the structures in position space, while e.g. \cite{Alday:2019nin} has explored at one-loop order the Mellin representation that was so fruitful at tree-level.

Further progress was made in Mellin space including string corrections for the simplest example of half-BPS correlators, the $\fourtwo$ correlator~\cite{Alday:2018kkw}.\footnote{A first generalisation to the one-loop supergravity Mellin amplitude for the $\langle\cO_2\cO_2\cO_p\cO_p\rangle$ family of correlators was made in~\cite{Alday:2019nin}, complementing the general position space approach of~\cite{Aprile:2019rep}.} However, an explicit position space representation for the string corrected one-loop amplitude is still missing, which is the purpose of this paper. In order to find such a representation we return to the position space bootstrap and conjecture the form of the position space amplitude. To this end we introduce a basis of polylogarithmic functions, many of which are familiar from the position space representations of the one-loop supergravity amplitudes. In some sense the string corrections to the one loop amplitudes are even simpler than the supergravity contributions, as the transcendental weight required is actually lower. The supergravity amplitudes require functions up to weight four while the string corrections (essentially due to the finite spin support of the spectrum) require only weights up to three. However, we find that we necessarily need a new ingredient, a weight three function $f^{(3)}$ with a more general set of singularities (or `letters'). While we derive explicit results only for the first few orders in the $1/\lambda$ expansion, the form of our one-loop bootstrap ansatz is in fact valid to all orders.

We are also able to make direct contact with the Mellin space results of \cite{Alday:2018kkw}, writing the spacetime expressions in a form in which they can be immediately expanded about the limit of the light-like square where the cross-ratios $u$ and $v$ are both small. In this way we confirm our position space results are consistent with the order $(\alpha')^3$ results of \cite{Alday:2018kkw}. We then provide new results in spacetime and in Mellin space for higher orders.

\subsection{The $\fourtwo$ correlator}
We are interested in the four-point correlator of the stress-tensor superprimary, which is a superconformal half-BPS operator of conformal dimension two, given by
\begin{align}
	\cO_2 = y^i y^j ~\text{Tr}\big(\Phi_i\Phi_j\big),
\end{align}
where $y^i$ is an auxiliary $so(6)$ vector obeying the null condition $y\cdot y=0$, such that $\cO_2$ is in the traceless symmetric representation $[0,2,0]$ of $su(4)$. This operator is dual to the graviton supermultiplet, whereas its higher charge versions $\cO_p$ are dual to a tower of Kaluza-Klein modes which arise from the ten-dimensional graviton upon compactification on the S$^5$ factor of the AdS$_5\times$S$^5$ background.

We consider the four-point function $\fourtwo$, which is constrained by superconformal symmetry to take the form~\cite{Eden:2000bk,Nirschl:2004pa}
\begin{align}\label{eq:partial_non-ren}
	\fourtwo = \fourtwo_{\text{free}}+g_{12}^2g_{34}^2~\mathcal{I}~\mathcal{H}(u,v),
\end{align}
where the propagator factors $g_{ij}=y_{ij}^2/x_{ij}^2$ (with $y_{ij}^2=y_i\cdot y_j$) carry the conformal weight and the scaling weights $y_i$ of the correlator. For convenience, we divide the correlator by a factor of $(N^2-1)^2$ such that its free part is then given by the following crossing symmetric combination of six propagator structures:
\begin{align}
\begin{split}
	\fourtwo_{\text{free}} &= 4\big(g_{12}^2g_{34}^2 + g_{13}^2g_{24}^2 + g_{14}^2g_{23}^2\big)\\ &~~~+16a\big(g_{12}g_{13}g_{24}g_{34}+g_{12}g_{14}g_{23}g_{34}+g_{13}g_{14}g_{23}g_{24}\big),
\end{split}
\end{align}
where we introduced the factor $a=1/(N^2-1)$. It is helpful to define the conformal and $su(4)$ R-symmetry cross-ratios by
\begin{equation}\label{eq:crossratios}
\begin{aligned}
	u= x \xb &= \frac{x_{12}^2x_{34}^2}{x_{13}^2 x_{24}^2},\qquad &&~v=(1-x)(1-\xb)=  \frac{x_{14}^2x_{23}^2}{x_{13}^2 x_{24}^2}, \\
	\frac{1}{\sigma}=y \bar y &= \frac{y_{12}^2 y_{34}^2}{y_{13}^2 y_{24}^2},\qquad &&\frac{\tau}{\sigma}=(1-y)(1-\bar y)=\frac{y_{14}^2 y_{23}^2}{y_{13}^2 y_{24}^2},
\end{aligned}
\end{equation}
which can be readily used to rewrite the free theory correlator as
\begin{align}\label{eq:free_2222}
	\fourtwo_{\text{free}} = 4g_{12}^2g_{34}^2\left(\Big[1+u^2\sigma^2+\frac{u^2\tau^2}{v^2}\Big]+4a\Big[u\sigma+\frac{u\tau}{v}+\frac{u^2\sigma\tau}{v}\Big]\right).
\end{align}
We call the remaining contribution to equation~\eqref{eq:partial_non-ren} the interacting part. Its dependence on the $su(4)$ variables $y$ and $\bar{y}$ is completely fixed by the superconformal Ward identities, taking the factorised form shown above with the factor $\mathcal{I}$ given by
\begin{align}
	\mathcal{I} = \frac{(x-y)(x-\bar{y})(\xb-y)(\xb-\bar{y})}{(y\bar{y})^2}.
\end{align}
The function $\mathcal{H}$ is then independent of the internal variables, which is a feature of the $\fourtwo$ correlator.\footnote{More generally, it is true for next-to-next-to-extremal correlators.} Furthermore, it is the only piece of the correlator which depends on the gauge coupling $g_{\text{YM}}$, thus containing all the non-trivial dynamics of the theory. In order to respect the full crossing symmetry of the correlator, $\mathcal{H}$ obeys the crossing transformations
\begin{align}\label{eq:crossing_H}
	\mathcal{H}(u,v) = \frac{1}{v^2}\mathcal{H}(u/v,1/v) = \frac{u^2}{v^2}\mathcal{H}(v,u).
\end{align}
This fact will be central to our bootstrap approach discussed later, as it places strong constraints on the functional form of $\mathcal{H}$. We will study this correlator in the supergravity limit, where one first takes the large $N$ limit with the 't Hooft coupling $\lambda=g_{\text{YM}}^2N$ fixed, and then expands each term around large $\lambda$. In this limit, the double expansion of $\mathcal{H}$ takes the form\footnote{\label{footnote}Note that at order $a^2$ we have kept only the genuine one-loop contributions which are induced by the presence of the tree-level terms $\mathcal{H}^{(1,k)}$. In particular, we have omitted the super-leading term $\l^{\frac{1}{2}}\mathcal{H}^{(2,-1)}$, corresponding to a quadratic divergence in the one-loop supergravity amplitude which is regulated by an $\mathcal{R}^4$ one-loop counterterm in string-theory. Furthermore, there are the additional terms $\l^{\frac{1}{2}}\mathcal{H}^{(2,1)}$ and $\l^{-1}\mathcal{H}^{(2,2)}$, which precede the genuine one-loop string correction $\ll\mathcal{H}^{(2,3)}$. These two terms correspond to the genus-one contributions to the modular completions of the tree-level terms $\lll\mathcal{H}^{(1,5)}$ and $\l^{-3}\mathcal{H}^{(1,6)}$, respectively. See references~\cite{Green:1999pv,Chester:2019pvm} for a more detailed discussion of these contributions.}
\begin{align}\label{eq:double_expansion}
\begin{split}
	\mathcal{H} &= ~~\,a \left( \mathcal{H}^{(1,0)}+\ll\mathcal{H}^{(1,3)}+\lll\mathcal{H}^{(1,5)}+\ldots\right)\\
				&~\,+ a^2 \left( \mathcal{H}^{(2,0)}+\ll\mathcal{H}^{(2,3)}+\lll\mathcal{H}^{(2,5)}+\ldots\right)+O(a^3),
\end{split}
\end{align}
where we use $a=1/(N^2-1)$ as our large $N$ expansion parameter. The term of order $a^0$ in equation~\eqref{eq:free_2222} is the contribution from disconnected free field theory. At first order in $a$, it receives tree-level corrections from supergravity, with $\mathcal{H}^{(1,0)}$ denoting the well known supergravity result~\cite{Arutyunov:2000py,Dolan:2001tt}, followed by an infinite tower of $1/\lambda$ suppressed string corrections $\mathcal{H}^{(1,k)}$. The structure of this $1/\lambda$ expansion is related to the low-energy expansion of the tree-level type IIB string amplitude in ten dimensions, the so called Virasoro-Shapiro amplitude, via the flat-space limit of the Mellin amplitudes corresponding to $\mathcal{H}^{(1,k)}$~\cite{Penedones:2010ue,Fitzpatrick:2011hu}. In other words, the $1/\lambda$ expansion arises from contact interaction vertices in the string theory effective action, where the order $\ll$ and $\lll$ terms descend from dimensional reduction of the $\mathcal{R}^4$ and $\partial^4\mathcal{R}^4$ supervertices, respectively. For higher corrections with $k\geq5$, $\lambda^{-\frac{k}{2}}$ corresponds to the $\partial^{2k-6}\mathcal{R}^4$ vertex, such that $\mathcal{H}^{(1,k)}$ is non-zero for $k\in\{0,3\}\cup\{5,6,7,\ldots\}$. The flat space limit was used in~\cite{Goncalves:2014ffa} to determine the first two string corrections, i.e. for $k=3,5$. These results were recently generalised to correlators of the form $\langle\cO_2\cO_2\cO_p\cO_p\rangle$ by using the bulk-point limit~\cite{Alday:2018pdi} and supersymmetric localisation~\cite{Binder:2019jwn}, and at order $\ll$ to correlators with arbitrary external charges $\langle\cO_p\cO_q\cO_r\cO_s\rangle$~\cite{Drummond:2019odu}.

Here we will be interested in the order $a^2$ terms of the expansion~\eqref{eq:double_expansion}, which correspond to one-loop amplitudes in AdS$_5$. The one-loop supergravity amplitude $\mathcal{H}^{(2,0)}$ was derived in~\cite{Aprile:2017bgs}, while its Mellin space representation was obtained recently in~\cite{Alday:2018kkw}. Moreover, the structure of string corrections at one-loop was addressed in that work in Mellin space, as we will briefly review in Section~\ref{sec:Mellin_space}. The purpose of this paper is to construct the explicit position space expressions $\mathcal{H}^{(2,k)}$, which we discuss in detail in Section~\ref{sec:position_space}.

Lastly, the unprotected (long) part of the $\fourtwo$ correlator, which receives contributions from both the free theory and the interacting part, admits a decomposition into superconformal blocks. In this case, the block expansion reads
\begin{align}\label{eq:block_deco}
	\fourtwo_{\text{long}} = g_{12}^2g_{34}^2~\mathcal{I}~\sum_{t,\ell}A_{t,\ell}\,G_{t,\ell}(\x,\xb),
\end{align}
where the sum runs over all long superconformal primaries with half-twist $t\equiv(\Delta-\ell)/2$ and spin $\ell$ which are present in the $su(4)$ singlet representation in the OPE of $\cO_2\times\cO_2$. The $A_{t,\ell}$ are squared three-point functions, $G_{t,\ell}$ is related to the four-dimensional conformal block and is given by~\cite{Dolan:2000ut,Dolan:2003hv}
\begin{align}\label{eq:conformal_block}
	G_{t,\ell}(\x,\xb) = (-1)^\ell (\x\xb)^t~\frac{\x^{\ell+1}F_{t+\ell+2}(\x)F_{t+1}(\xb)-\xb^{\ell+1}F_{t+\ell+2}(\xb)F_{t+1}(\x)}{\x-\xb},
\end{align}
with $F_{\rho}(\x)={}_2F_1\left(\rho,\rho,2\rho;\x\right)$ being the standard hypergeometric function.

In the double expansion~\eqref{eq:double_expansion} around the supergravity limit, the leading contribution to the spectrum of exchanged operators is given by unprotected double-trace operators of classical dimension $\Delta^{(0)}=2t+\ell$ and even spin $\ell$. Generically, there are many such operators with the same quantum numbers, leading to a mixing problem. In the singlet channel and for a given half-twist $t$, there are $t-1$ such operators of the form
\begin{align}\label{eq:degenerate_operators}
 \cO_2\square^{t-2}\partial^\ell\cO_2,~\cO_3\square^{t-3}\partial^\ell\cO_3,~\ldots~,~\cO_t\square^0\partial^\ell\cO_t,
\end{align}
which we label by $i=1,\ldots,t-1$. Their dimensions admit the expansion
\begin{align}\label{eq:def_anom_dims}
\begin{split}
	\Delta_{t,\ell} = \Delta^{(0)} + 2a&\left(\eta^{(1,0)}_i + \ll\eta^{(1,3)}_i + \lll\eta^{(1,5)}_i + \ldots\right)\\
								  +2a^2&\left(\eta^{(2,0)}_i + \ll\eta^{(2,3)}_i + \lll\eta^{(2,5)}_i + \ldots\right)+O(a^3).
\end{split}
\end{align}
The above mixing problem has been addressed at leading order, resulting in explicit formulae for their leading order three-point functions $A_{t,\ell,i}^{(0)}$ and a compact formula for their supergravity anomalous dimensions $\eta^{(1,0)}_i$~\cite{Aprile:2017xsp,Aprile:2018efk}. Further $1/\lambda$ corrections to the spectrum of double-trace operators have been addressed in~\cite{Drummond:2019odu}, revealing surprisingly simple patterns in their string corrected anomalous dimensions $\eta^{(1,3)}_i$ and $\eta^{(1,5)}_i$.
\section{Bootstrap method in position space}\label{sec:position_space}\setcounter{equation}{0}
We begin by reviewing how to obtain the so called double discontinuity (i.e. the $\log^2u$-piece of the correlator) from tree-level data only. The double discontinuity, in particular in the context of holographic correlators, has recently received a lot of attention. Namely, it is the central object from which one-loop OPE data can be extracted without first constructing the full one-loop correlator~\cite{Caron-Huot:2017vep}. As we discuss in Section~\ref{sec:bootstrap_problem}, this then allows us to pose a well-defined bootstrap problem, whose solution completely determines the one-loop correlators $\mathcal{H}^{(2,k)}$ from a given double discontinuity $\mathcal{H}^{(2,k)}|_{\log^2u}$, up to a finite number of well understood ambiguities. Notably, a new ingredient enters our ansatz of transcendental functions: it turns out that a certain function of transcendental weight three with a new type of singularity has to be included. We describe this new ingredient in Section~\ref{sec:f3}.
\subsection{String corrected double discontinuities}\label{sec:double_disc}
Let us start by discussing the specific form of the double discontinuities which arise in the $1/\lambda$ expansion of the one-loop correlator at order $a^2$. Crucially, the double discontinuity is fully determined by tree-level data through the conformal block decomposition~\eqref{eq:block_deco}. More explicitly, we can compute the $\log^2u$ part of $\mathcal{H}^{(2,k)}$ from spectral data at order $a$:\footnote{Note that for $m\neq n$ we need to include the two identical contributions $\mathcal{D}^{m|n}$ and $\mathcal{D}^{n|m}$.}
\begin{align}
	\left.\mathcal{H}^{(2,k)}(x,\xb)\right|_{\log^2(u)} = \sum_{m+n=k}\mathcal{D}^{m|n}(\x,\xb),
\end{align}
where we have introduced the notation
\begin{align}\label{eq:double_disc_sum}
	\mathcal{D}^{m|n}(\x,\xb)=\frac{1}{2}\sum_{t,\ell}\sum_{i=1}^{t-1}A_{t,\ell,i}^{(0)}~\eta_i^{(1,m)}\eta_i^{(1,n)}~G_{t+2,\ell}(x,\xb),
\end{align}
with $\eta^{(1,m)}_i$ being the tree-level anomalous dimensions at order $\lambda^{-\frac{m}{2}}$, and $i$ labelling the set of exchanged double-trace operators~\eqref{eq:degenerate_operators}. Recalling equation~\eqref{eq:double_expansion} and the discussion below, the general structure of the $1/\lambda$ expansion at order $a$ demands that the integers $k,m,n$ in the above equation are drawn from the set $\{0,3\}\cup\{5,6,7,8\ldots\}$, with the constraint $m+n=k$. Note that when $k$ is large enough to accommodate for different partitions into $(m,n)$, we get more than one contribution to the double discontinuity at that order in $1/\lambda$.\footnote{The first instance of this happens already for $k=6$ at order $\lambda^{-3}$, for which there are the two distinct possibilities $(m,n)=(0,6)$ or $(3,3)$. These contributions correspond to the insertions of $\sg$$|$$\partial^6\mathcal{R}^4$ and $\mathcal{R}^4$$|$$\mathcal{R}^4$ vertices, respectively.} See also Table~\ref{tab:terms_at_one-loop} for the first few one-loop terms in the $1/\lambda$ expansion. 

The supergravity double discontinuity $\mathcal{D}^{0|0}$ was explicitly computed in~\cite{Aprile:2017bgs}, where it was used to reconstruct the full one-loop function $\mathcal{H}^{(2,0)}$.\footnote{The supergravity double discontinuity can also be obtained by ``squaring'' tree-level correlators of the form $\langle\cO_2\cO_2\cO_p\cO_p\rangle$~\cite{Alday:2017xua,Alday:2017vkk}, without solving the mixing-problem explicitly. Similarly, string corrected double discontinuities at a given order in $1/\lambda$ can be obtained from the $\langle\cO_2\cO_2\cO_p\cO_p\rangle$ family of correlators up to that order, without explicit reference to the corresponding string anomalous dimensions~\cite{Alday:2018pdi}.} In contrast to the supergravity case however, it turns out that the string corrected double discontinuities ($\mathcal{D}^{m|n}$ with $m,n\neq0$) resum into expressions of up to transcendental weight one only, compared to up to weight two terms in supergravity. The reason for this is the spin truncation in the string corrected spectrum. To be explicit, these double discontinuities are of the form
\begin{align}\label{eq:double_discs}
	\mathcal{D}^{m|n}(\x,\xb)=u^2\left(\frac{p_1^{m|n}(x,\xb)}{(x-\xb)^{q-1}}+\frac{p_2^{m|n}(x,\xb)\left(\log(1-x)-\log(1-\xb)\right)}{(x-\xb)^{q}}\right),
\end{align}
where the odd denominator power is given by $q=2(m+n)+15$ and $p_1$, $p_2$ are symmetric polynomials in $(\x,\xb)$ of the same degree as their denominator. This simple structure for the double discontinuities was already obtained in~\cite{Alday:2018pdi}\footnote{In this reference the authors also propose a basis of special functions $S_{\ell}^{(q)}$, into which all string double discontinuities $\mathcal{D}^{m|n}$ can be decomposed.}, and we find complete agreement with their results by performing the sum~\eqref{eq:double_disc_sum} for different cases.

Note that the double discontinuities have a symmetry under the $1\leftrightarrow2$ crossing transformation, which acts on the cross-ratios as $\x\rightarrow\x'\equiv x/(x-1)$, and similarly for $\xb$. This symmetry is inherited from the full crossing symmetry of the $\fourtwo$ correlator as it survives the s-channel OPE decomposition~\eqref{eq:block_deco}. As a formula, we have
\begin{align}
	\mathcal{D}^{m|n}(\x',\xb')=v^2\, \mathcal{D}^{m|n}(\x,\xb).
\end{align}

In the following, we provide an algorithm on how to uplift the double discontinuity $\mathcal{D}^{m|n}$ to the corresponding fully crossing symmetric function $\mathcal{H}^{(2,k)}(u,v)$.
\subsection{The bootstrap problem}\label{sec:bootstrap_problem}
In order to simplify the crossing transformations~\eqref{eq:crossing_H} of the interacting part $\mathcal{H}(u,v)$, we introduce an auxiliary function $\mathcal{F}$ by
\begin{align}\label{eq:F_definition}
	\mathcal{F}(u,v) = \frac{(\x-\xb)^4}{u^2}\,\mathcal{H}(u,v),
\end{align}
such that $\mathcal{F}(u,v)$ transforms without picking up any prefactors under crossing:
\begin{align}\label{eq:crossing_F}
	\mathcal{F}(u,v) = \mathcal{F}(u/v,1/v) = \mathcal{F}(v,u).
\end{align}
Evidently, $\mathcal{F}$ inherits an analogous double expansion as $\mathcal{H}$ in~\eqref{eq:double_expansion}. Guided by the explicit form of the double discontinuities as given in~\eqref{eq:double_discs}, we propose the following structure for the functions $\mathcal{F}^{(2,k)}$ ($k>0$):
\begin{align}\label{eq:ansatz_F}
\begin{split}
	\mathcal{F}^{(2,k)}(u,v)&=A_1(\x,\xb)f^{(3)}(\x,\xb)+\big(A_2(\x,\xb)\log u-A_2(1-\x,1-\xb)\log v\big)\phi^{(1)}\left(x,\xb\right)\\
					&~~~+A_3(\x,\xb)\log^2 u + A_3\Big(\frac{x-1}{x},\frac{\xb-1}{\xb}\Big)\log^2\frac{u}{v} + A_3\Big(\frac{1}{1-x},\frac{1}{1-\xb}\Big)\log^2 v\\
					&~~~+A_4(\x,\xb)\phi^{(1)} \left(x,\xb\right)+\left(A_5(\x,\xb)\log u + A_5(1-\x,1-\xb)\log v\right)+A_6(\x,\xb).
\end{split}
\end{align}
The six coefficient functions $A_i(\x,\xb)$ are constrained by crossing symmetry and the explicit form of the double discontinuities $\mathcal{D}^{m|n}$, see equation~\eqref{eq:double_discs}, to be of the general form
\begin{align}\label{eq:ansatz_A_poly}
	A_i(\x,\xb) = \frac{1}{(\x-\xb)^d}~\sum_{r=0}^{d}\sum_{s=0}^{d-r}\, a^{i}_{r,s}\,u^r v^s,
\end{align}
with a finite number of free parameters $a^{i}_{r,s}$ and fixed denominator powers $d=2k+11$ ($d=2k+10$) in case $A_i$ is antisymmetric (symmetric) under $\x\leftrightarrow\xb$, see below. Recall the relation $k=m+n$, where $m$ and $n$ label the double discontinuity $\mathcal{D}^{m|n}$ at order $\lambda^{-\frac{k}{2}}$.\footnote{Note that the difference between the denominator powers $d$ here and $q$ in~\eqref{eq:double_discs} is due to the explicit $(x-\xb)^4$ factor in the definition of $\mathcal{F}$, according to equation~\eqref{eq:F_definition}.}

In order to ensure the exchange symmetry $\x\leftrightarrow\xb$ and the full crossing symmetry~\eqref{eq:crossing_F} of $\mathcal{F}^{(2,k)}(u,v)$, the coefficient functions $A_i$ obey the following relations:
\begin{equation}\label{eq:crossing_A_i}
\begin{alignedat}{2}
	A_1(\x,\xb) &=-A_1(\xb,\x),~~~~~~~&&A_1(\x,\xb)=-A_1\left(\frac{1}{\x},\frac{1}{\xb}\right)=-A_1(1-\x,1-\xb),\\
	A_2(\x,\xb) &=-A_2(\xb,\x),&&A_2(\x,\xb)=-A_2\left(\frac{x}{x-1},\frac{\xb}{\xb-1}\right),\\
	A_3(\x,\xb) &=A_3(\xb,\x),&&A_3(\x,\xb)=A_3\left(\frac{1}{\x},\frac{1}{\xb}\right),\\
	A_4(\x,\xb) &=-A_4(\xb,\x),&&A_4(\x,\xb)=-A_4\left(\frac{1}{\x},\frac{1}{\xb}\right)=-A_4(1-\x,1-\xb),\\
	A_5(\x,\xb) &=A_5(\xb,\x),&&A_5(\x,\xb)=A_5\left(\frac{x}{x-1},\frac{\xb}{\xb-1}\right),\\
	A_6(\x,\xb) &=A_6(\xb,\x),&&A_6(\x,\xb)=A_6\left(\frac{1}{\x},\frac{1}{\xb}\right)=A_6(1-\x,1-\xb),
\end{alignedat}
\end{equation}
and each of the coefficient functions $A_2$ and $A_5$ obey the additional constraint:
\begin{align}\label{eq:additional_A_constraints}
\begin{split}
	A_2(\x,\xb)+A_2\left(\frac{1}{1-\x},\frac{1}{1-\xb}\right)-A_2(1-\x,1-\xb) &= 0,\\
	A_5(\x,\xb)+A_5\left(\frac{1}{1-\x},\frac{1}{1-\xb}\right)+A_5(1-\x,1-\xb) &= 0.
\end{split}
\end{align}

The main new feature of the ansatz presented in~\eqref{eq:ansatz_F} is the presence of $f^{(3)}(x,\xb)$, which is an antisymmetric single-valued function of transcendental weight three. As this function is new in the context of AdS amplitudes, involving a new type of singularity compared to the previously known supergravity case, we will describe it in more detail in the next section (see also Appendix~\ref{App-f3}).

On the other hand, the function $\phi^{(1)}(x,\xb)$ is the well-known one-loop massless box-integral in four-dimensions.\footnote{$\phi^{(1)}(x,\xb)$ is the first member of the more general series of ladder functions $\phi^{(l)}(x,\xb)$~\cite{Usyukina:1992jd}.} It is an antisymmetric weight-two function given by
\begin{align}\label{eq:phi1}
	\phi^{(1)}(x,\xb) = 2\big(\text{Li}_2(x)-\text{Li}_2(\xb)\big)+\log(u)\big(\log(1-x)-\log(1-\xb)\big),
\end{align}
and obeys the symmetries
\begin{align}
	\phi^{(1)}(x,\xb) = -\phi^{(1)}(\xb,x) = -\phi^{(1)}(1-x,1-\xb) =-\phi^{(1)}(1/x,1/\xb).
\end{align}
Note that a very similar type of ansatz in terms of ladder functions was used before to bootstrap one-loop supergravity contributions to various correlators, see~\cite{Aprile:2017bgs,Aprile:2017qoy,Aprile:2019rep}. Let us highlight the two main differences of our ansatz~\eqref{eq:ansatz_F} to the one-loop supergravity case:
\begin{itemize}
\item The ansatz for $\mathcal{F}^{(2,k)}(u,v)$ presented above has maximal transcendental weight three, compared to up to weight-four contributions in supergravity. This difference is ultimately a consequence of the spin truncation of the string corrected spectrum. A truncation to finite spin produces resummed double discontinuities of the form depicted in~\eqref{eq:double_discs}, which has terms of maximal weight one. In contrast, the supergravity spectrum has infinite spin support, resulting in up to weight-two contributions in the corresponding double discontinuity.

\item As mentioned before, the presence of the function $f^{(3)}(x,\xb)$ is a novelty in the context of AdS amplitudes. However, one can already see from the structure of the double discontinuities $\mathcal{D}^{m|n}$ that a new ingredient is required: as we will discuss shortly, an ansatz with ladder functions only would enforce a structure on the $p_2^{m|n}$ polynomial of $\mathcal{D}^{m|n}$ which is not observed from direct resummations. We are therefore led to conclude that we need a new contribution in our ansatz, which we denote by $f^{(3)}(x,\xb)$ and whose full characterisation we postpone to Section~\ref{sec:f3}.
\end{itemize}
This completes the description of our ansatz for the one-loop string amplitudes $\mathcal{F}^{(2,k)}$. Next, we continue by describing the conditions we impose in order to constrain the free parameters $a^{i}_{r,s}$ in the ansatz~\eqref{eq:ansatz_F} described above.
\subsubsection*{Constraining the free parameters}
In analogy to the position space bootstrap method for one-loop supergravity correlators~\cite{Aprile:2017bgs,Aprile:2017qoy,Aprile:2019rep}, there are two steps in constraining the free parameters $a^{i}_{r,s}$ in our ansatz:
\begin{enumerate}
\item \textbf{Matching the double discontinuity:}\\
	  The contribution of our ansatz~\eqref{eq:ansatz_F} to the $\log^2u$ term is given by
	  \begin{align}\label{eq:double_disc_F}
	  \begin{split}
	  		\mathcal{F}^{(2,k)}(u,v)|_{\log^2u} =& \left(-\frac{1}{2}A_1(\x,\xb)+A_2(\x,\xb)\right)\big(\log(1-\x)-\log(1-\xb)\big)\\
	  		&~~ +A_3(\x,\xb)+A_3\Big(\frac{x-1}{x},\frac{\xb-1}{\xb}\Big).
	  \end{split}
	  \end{align}
	  Matching this against the corresponding double discontinuity $\mathcal{D}^{m|n}$ fully fixes the coefficient functions $A_i(\x,\xb)$ for $i=1,2,3$.
	  
	  It is a fact that the polynomials $p_2^{m|n}$ in the resummed double discontinuities do not obey the first line of~\eqref{eq:additional_A_constraints}, and hence we require a non-zero contribution from the new weight-three function $f^{(3)}(\x,\xb)$ with coefficient $A_1(\x,\xb)$.
\item \textbf{Pole cancellation:}\\
	  The ansatz for the function $\mathcal{H}^{(2,k)}=\frac{u^2}{(\x-\xb)^4}\mathcal{F}^{(2,k)}$ contains explicit denominator factors, potentially giving rise to up to $q$ poles at $x=\xb$. Demanding that the full function $\mathcal{H}^{(2,k)}$ is free from such unphysical poles is what we mean by pole cancellation. Concretely, by imposing as many zeroes between the functions in the numerator of $\mathcal{F}^{(2,k)}$ as there are poles we find further non-trivial constraints amongst the remaining free parameters in $A_4(\x,\xb)$, $A_5(\x,\xb)$ and $A_6(\x,\xb)$.
\end{enumerate}
Carrying out the above two steps yields a definite answer for $\mathcal{H}^{(2,k)}$, and we are left with only a small number of remaining free parameters. We call these functions which pass all of the above constraints, and whose coefficients we therefore are not able to determine, ambiguities.

By construction, the ambiguities do not contribute to the double discontinuity, are fully crossing symmetric by themselves and free of unphysical poles.\footnote{In the language of~\cite{Aprile:2019rep}, we may call the ambiguities to be ``tree-like'' in the sense that they are of the form of $\dbar{}$-functions.} They are given by (linear combinations of) $\dbar{}$-functions with their $(\x-\xb)$ denominator power bounded by the corresponding denominator in the ansatz for $A_4(\x,\xb)$ in~\eqref{eq:ansatz_A_poly}: $d=2k+11$. We find that the ambiguities have finite spin support, and hence they are most conveniently described in Mellin space because their Mellin amplitudes are only polynomial. We therefore postpone the general discussion of ambiguities to Section~\ref{sec:ambiguities}, where we give a full classification in terms of polynomial Mellin amplitudes. For now, we simply list the total number of ambiguities and their maximal spin contributions $\ell_{\text{max}}$ for the first couple of orders in the $1/\lambda$ expansion in Table~\ref{tab:terms_at_one-loop}.
\renewcommand{\arraystretch}{1.3}
\begin{table} 
\begin{center}
\begin{tabular}{c|c|c|c|c}
$1/\lambda$	order		&corresponding supervertices&$\frac{1}{(\x-\xb)^{q}}$&$\namb$&$\ell_{\text{max}}$\\\hline
1						&$\sg$$|$$\sg$&15&1&0\\
$\lambda^{-\frac{3}{2}}$&$\sg$$|$$\mathcal{R}^4$&21&4&4\\
$\lambda^{-\frac{5}{2}}$&$\sg$$|$$\partial^4\mathcal{R}^4$&25&7&6\\
$\lambda^{-3}$			&$\sg$$|$$\partial^6\mathcal{R}^4$,~$\mathcal{R}^4$$|$$\mathcal{R}^4$&27&8&6\\
$\lambda^{-\frac{7}{2}}$&$\sg$$|$$\partial^8\mathcal{R}^4$&29&10&8\\
$\lambda^{-4}$			&$\sg$$|$$\partial^{10}\mathcal{R}^4$,~$\mathcal{R}^4$$|$$\partial^4\mathcal{R}^4$&31&12&8\\
$\lambda^{-\frac{9}{2}}$&$\sg$$|$$\partial^{12}\mathcal{R}^4$,~$\mathcal{R}^4$$|$$\partial^6\mathcal{R}^4$&33&14&10\\
$\lambda^{-5}$			&$\sg$$|$$\partial^{14}\mathcal{R}^4$,~$\mathcal{R}^4$$|$$\partial^8\mathcal{R}^4$,~$\partial^4\mathcal{R}^4$$|$$\partial^4\mathcal{R}^4$&35&16&10\\
\end{tabular}
\end{center}
\caption{List of one-loop terms in the $1/\lambda$ expansion and their corresponding vertices in the effective string theory action, where $\sg$ stands for an insertion of the supergravity anomalous dimension. We give the denominator powers $q=2k+15$ of the spacetime functions $\mathcal{H}^{(2,k)}$, the total number of ambiguities $\namb$ as well as their maximal spin support $\ell_{\text{max}}$. Note that in general there can be more than one term contributing to the same order in $1/\lambda$, the first occurrence of this happening at order $\lambda^{-3}$.}
\label{tab:terms_at_one-loop}
\end{table}
\subsubsection*{Results}
Following the bootstrap construction outlined above, we explicitly computed the full one-loop amplitudes $\mathcal{H}^{(2,3)}$ and $\mathcal{H}^{(2,5)}$, as well as the contributions to $\mathcal{H}^{(2,6)}$, $\mathcal{H}^{(2,8)}$ and $\mathcal{H}^{(2,10)}$ which descend from the double discontinuities $\mathcal{D}^{3|3}$, $\mathcal{D}^{3|5}$ and $\mathcal{D}^{5|5}$, respectively. In all cases the results are in agreement with the general patterns described in Table~\ref{tab:terms_at_one-loop}. For the amplitudes $\mathcal{H}^{(2,k)}$ with $k=3,5,6$, we attach the corresponding lists of polynomials $A_i(\x,\xb)$ in an ancillary file to the \texttt{arXiv} submission.

From these position space results, one can then extract further subleading spectral data. However, as a consequence of the degeneracy in the double-trace spectrum discussed around equation~\eqref{eq:degenerate_operators}, this is possible only for the lowest twist contribution, where there is a single operator. We computed the order $a^2\ll$ and $a^2\lll$ one-loop anomalous dimensions $\eta^{(2,3)}$ and $\eta^{(2,5)}$ at twist four, finding agreement with the results of~\cite{Binder:2019jwn}.

\subsection{A new ingredient: $f^{(3)}$}\label{sec:f3}
In finding a suitable crossing symmetric function which matches the double discontinuity, we encountered the need to include functions beyond the ladder class encountered at one loop in the supergravity results described in \cite{Aprile:2017bgs,Aprile:2017qoy,Aprile:2019rep}. In fact the new function is both simpler, in that it is of transcendental weight three only, and more complicated, as it involves new singularities (`letters') of the form $x-\bar{x}$ not found in the ladders. It is also single-valued in the same sense as the ladder functions e.g. $\phi^{(1)}(x,\bar{x})$ given in eq. (\ref{eq:phi1}).

We may characterise $f^{(3)}$ by its total derivative,
\begin{align}\label{f3totald}
df^{(3)}(x,\bar{x}) = \quad &\bigl[- 2 \phi^{(1)}(x,\bar{x}) +\tfrac{1}{2}\log^2 v - \log u \log v \bigr] d \log x  \notag \\
+ &\bigl[- 2 \phi^{(1)}(x,\bar{x}) -\tfrac{1}{2}\log^2 v + \log u \log v \bigr] d\log \bar{x}  \notag \\
+ & \bigl[- 2 \phi^{(1)}(x,\bar{x}) -\tfrac{1}{2}\log^2 u + \log u \log v \bigr] d \log (1-x) \notag \\
+ & \bigl[- 2 \phi^{(1)}(x,\bar{x}) +\tfrac{1}{2}\log^2 u - \log u \log v \bigr] d \log (1-\bar{x}) \notag \\
+ & \bigl[6\phi^{(1)}(x,\bar{x})\bigr] d \log (x-\bar{x}) \,,
\end{align}
together with its symmetry property, 
\be
f^{(3)}(x,\bar{x}) =  - f^{(3)}(\bar{x},x)
\ee
which implies $f^{(3)}(x,x)=0$. It also obeys antisymmetry under the crossing transformations
\be
f^{(3)}(1-x,1-\bar{x}) = - f^{(3)}(x,\bar{x}) = f^{(3)}(1/x,1/\bar{x})\,.
\ee
Up to adding a linear combination of single-valued HPLs it can be identified with the weight three function called $\mathcal{Q}_3$ found in \cite{Chavez:2012kn}. Functions with the same type of singularities are also needed in perturbation theory to describe the correlators of half-BPS operators at three loops \cite{Drummond:2013nda,Chicherin:2015edu}.

We may make the $\log u$ discontinuities of $f^{(3)}(x,\bar{x})$ more transparent by writing
\be
f^{(3)}(x,\bar{x}) = \log^2 u \tilde{f}^{(1)}(x,\bar{x}) + \log u \tilde{f}^{(2)}(x,\bar{x}) + \tilde{f}^{(3)}(x,\bar{x})\,,
\label{f3loguform}
\ee
where the $\tilde{f}^{(k)}$ have no $\log u$ discontinuities.
The double $\log u$ discontinuity is given by
\be
\tilde{f}^{(1)}(x,\bar{x}) = -\frac{1}{2} \bigl[\log (1-x)- \log(1-\bar{x})\bigr] \,,
\ee
as already indicated in equation~\eqref{eq:double_disc_F}.
The single $\log u$ discontinuity can also be simply integrated to obtain
\begin{align}
\tilde{f}^{(2)}(x,\bar{x}) 
=&+6\, {\rm Li}_2\biggl(\frac{\bar{x}-x}{1-x}\biggr) + 2\bigl({\rm Li}_2(x) - {\rm Li}_2(\bar{x})\bigr) \notag \\
&+\frac{5}{2} \log^2(1-x) - 3 \log(1-x) \log(1-\bar{x}) + \frac{1}{2} \log^2(1-\bar{x})\,.
\label{f2texp}
\end{align}
The $\log^0 u$ term can be integrated in terms of hyperlogarithms (or Goncharov polylogs). We discuss this more in Appendix \ref{App-f3} where we also discuss various techniques for writing the function in a form suitable for comparison to Mellin representations of the one-loop string amplitude, which we address in the next section.

\section{Comparison with Mellin space}\label{sec:Mellin_space}\setcounter{equation}{0}
The Mellin space formalism~\cite{Mack:2009mi,Mack:2009gy,Penedones:2010ue,Fitzpatrick:2011ia} has turned out to be an efficient framework for constructing holographic correlators, as it makes manifest the analytic structure of the correlator. Considering the bootstrap problem for correlators at strong coupling in Mellin space has led to a wealth of new results at tree-level and more recently also at one-loop, both of which we briefly review in this section. In particular, we verify that our position space results for one-loop string amplitudes are in agreement with the Mellin space structures found in~\cite{Alday:2018kkw}, and we furthermore provide a number of new explicit Mellin amplitudes at higher orders in $1/\lambda$.

The Mellin representation of the interacting part is defined by the integral transform
\begin{align}\label{eq:mellintrafo}
	\mathcal{H}(u,v) = \int_{-i\infty}^{i\infty} \frac{ds}{2}\frac{dt}{2} ~u^{\frac{s}{2}}v^{\frac{t}{2}-2}~\me(s,t)~\Gamma^2\Big(\frac{4-s}{2}\Big)\Gamma^2\Big(\frac{4-t}{2}\Big)\Gamma^2\Big(\frac{4-\tilde{u}}{2}\Big),
\end{align}
where the Mellin variables $(s,t,\tilde{u})$ satisfy the constraint equation $s+t+\tilde{u}=4$. In order to obey the correct crossing transformations, the Mellin amplitude $\me(s,t)$ has the symmetries
\begin{align}\label{eq:crossing_M}
	\me(s,t)=\me(s,\tilde{u})=\me(t,s).
\end{align}
Furthermore, the Mellin amplitude inherits an analogous strong coupling expansion from $\mathcal{H}(u,v)$, see equation~\eqref{eq:double_expansion}. The only subtlety is that the long part of the free theory contribution has vanishing Mellin amplitude; it is instead recovered by the correct choice of integration contour, see~\cite{Rastelli:2017udc} for further details. Hence $\me(s,t)$ admits an expansion of the form
\begin{align}\label{eq:double_expansion_Mellin}
\begin{split}
	\me = ~~a&\left( \me^{(1,0)}+\ll\me^{(1,3)}+\lll\me^{(1,5)}+\ldots\right)\\
	      +a^2&\left(\me^{(2,0)}+\ll\me^{(2,3)}+\lll\me^{(2,5)}+\ldots\right)+O(a^3),
\end{split}
\end{align}
where at order $a^2$ we again keep only the genuine one-loop terms.\footnote{See footnote~\ref{footnote}.}

We start by first reviewing the tree-level terms $\me^{(1,k)}$, which will then be useful in Section~\ref{sec:ambiguities} for the discussion of the ambiguities our position space bootstrap method is not able to fix. Lastly, we describe the general structure of the one-loop Mellin amplitudes $\me^{(2,k)}$, providing a new result for $\me^{(2,5)}$ and new partial results at orders $k=8,10$.
\subsection{Review of tree-level Mellin amplitudes}\label{sec:Mellin_tree_level}
Tree-level Witten diagrams take a particularly simple form when represented in Mellin space. In the case of tree-level supergravity, they are rational functions of the Mellin variables with a prescribed set of poles, which correspond to exchanged single-trace operators in a particular Witten diagram. Beyond supergravity, further string corrections are even simpler in Mellin space, as their Mellin amplitudes are only polynomial. 

In our conventions, the Mellin amplitude of the well known tree-level supergravity result $\mathcal{H}^{(1,0)}$ is given by~\cite{Goncalves:2014ffa}
\begin{align}
	\me^{(1,0)} = \frac{128}{(s-2)(t-2)(\tilde{u}-2)},
\end{align}
which was then generalised to a compact formula for all tree-level four-point functions with arbitrary external charges~\cite{Rastelli:2016nze,Rastelli:2017udc}.

String corrections to the supergravity result descend from higher derivative interaction terms in the AdS$_5\times$S$^5$ effective action, which are of the schematic form $\partial^{2n}\mathcal{R}^4$. They give rise to contact Witten diagrams with spin truncated conformal block decompositions, and hence have polynomial Mellin amplitudes. A basis of polynomial Mellin amplitudes obeying the crossing equation~\eqref{eq:crossing_M} is given by the monomials $\sigma_2^p\sigma_3^q$, where $\sigma_n\equiv s^n+t^n+\tilde{u}^n$~\cite{Alday:2014tsa}. The conformal block decomposition of such a monomial has finite spin support, with only even spins $\ell\leq\ell_{\text{max}}=2(p+q)$ contributing. At a given order $\lambda^{-\frac{k}{2}}$ in the $1/\l$ expansion, the Mellin amplitude $\me^{(1,k)}$ is of the form
\begin{align}\label{eq:polynomial_ansatz}
	\sigma_2^p\sigma_3^q + \text{subleading terms},
\end{align}
with $p$, $q$ such that $k=3+2p+3q$. The first two string corrections (for $k=3,5$) have been addressed before using various methods~\cite{Goncalves:2014ffa,Alday:2018pdi,Binder:2019jwn}\footnote{In~\cite{Alday:2018pdi,Binder:2019jwn}, the authors consider more general correlators of the form $\langle\cO_2\cO_2\cO_p\cO_p\rangle$ at the first two orders $\ll$ and $\lll$. At order $\ll$, a generalisation to the correlator with arbitrary external charges was achieved in~\cite{Drummond:2019odu}.}, giving
\begin{align}
	\me^{(1,3)} = 1920\zeta_3,\qquad \me^{(1,5)} = \left(10080\sigma_2 - 30240\right)\zeta_5,
\end{align}
in our conventions.
\subsection{One-loop ambiguities}\label{sec:ambiguities}
Before discussing the structure of one-loop Mellin amplitudes, let us describe the ambiguities which are left unfixed by our position space bootstrap method. They are exactly of the form of tree-level string amplitudes which can be written in the basis of monomials $\sigma_2^p\sigma_3^q$ as discussed above, and hence we address them here.

At one-loop order, the structure of the $1/\lambda$ expansion follows from the low-energy expansion of the ten-dimensional genus-one IIB superstring amplitude~\cite{Green:1999pv}, leading to a different counting of the powers in $1/\lambda$: along with an additional factor of $a=1/(N^2-1)$, it is simply shifted by a power of $\lambda^{2}$ compared to the tree-level expansion.\footnote{In fact, this results in a constant Mellin amplitude contributing to a super-leading power at order $\lambda^{\frac{1}{2}}$, whose coefficient was fixed in~\cite{Chester:2019pvm}.} In other words, at one-loop order $a^2\lambda^{-\frac{k}{2}}$, one finds contributions of monomials $\sigma_2^p\sigma_3^q$ with $2p+3q\leq k+1$, in comparison to $2p+3q\leq k-3$ at tree-level.

In the supergravity case ($k=0$), there is a single one-loop ambiguity with constant Mellin amplitude, whose coefficient $\alpha$ was left unfixed in the result for $\mathcal{H}^{(2,0)}$ in~\cite{Aprile:2017bgs}. Recently, its value was determined by using supersymmetric localisation to be~\cite{Chester:2019pvm}
\begin{align}
	\alpha=60.
\end{align}

Finally, we fully characterise the one-loop ambiguities which arise in the string corrected correlators $\mathcal{H}^{(2,k)}$. According to the counting mentioned above, these ambiguities are enumerated by pairs of integers $(p,q)$ obeying the constraint $2p+3q\leq k+1$. By combining the above arguments, we can parametrise the set of ambiguities at any order $\lambda^{-\frac{k}{2}}$ by
\begin{align}
	\sum_{p,q\geq0}{\vphantom{\sum}}'~\alpha_{p,q}^{(k)}~\sigma_2^p\sigma_3^q,
\end{align}
where the primed sum is over $p$ and $q$ such that $2p+3q\leq k+1$. The total number of ambiguities $\namb(k)$ can be computed by expanding the generating function
\begin{align}
	\frac{1}{y(1-y)(1-y^2)(1-y^3)}-\frac{1}{y} = \sum_{k=0}^{\infty} \namb(k)y^k.
\end{align}
For the first couple of orders in the $1/\lambda$ expansion, we give the total number of ambiguities and their maximal spin support $\ell_{\text{max}}$ in Table~\ref{tab:terms_at_one-loop}.
\subsection{Mellin amplitudes at one-loop}\label{sec:Mellin_one_loop}
Let us now turn our attention to the structure of one-loop Mellin amplitudes for the $\fourtwo$ correlator, which has only recently been addressed for the first time in~\cite{Alday:2018kkw}.
The Mellin space approach is complementary to the bootstrap method in position space, as employed in~\cite{Aprile:2017bgs,Aprile:2017qoy,Aprile:2019rep} for one-loop supergravity, and  as outlined in the above Section~\ref{sec:position_space} for one-loop string corrections. We will give a short review of the known Mellin space results, both for one-loop supergravity and one-loop string corrections, and show agreement with our results in position space.
\subsubsection*{One-loop supergravity}
The one-loop supergravity Mellin amplitude is given in terms of the double infinite sum
\begin{align}
	\me^{(2,0)} = \sum_{m,n=2}^\infty \frac{c_{mn}}{(s-2m)(t-2n)} + \frac{c_{mn}}{(t-2m)(\tilde{u}-2n)} + \frac{c_{mn}}{(\tilde{u}-2m)(s-2n)},
\end{align}
 with constant coefficients $c_{mn}=c_{nm}$ given in equations (30) and (31) of~\cite{Alday:2018kkw}. Note that the simultaneous double poles in $s$ and $t$ in the above expression are necessary to reproduce the $\log^2(u)\log^2(v)$ part of the position space double discontinuitiy $\mathcal{D}^{(0|0)}$. Quite surprisingly, it turns out that any additional single poles are absent from $\me^{(2,0)}$, and therefore the entire Mellin amplitude can be fixed by matching against the double discontinuity $\mathcal{D}^{(0|0)}$. Recently, a generalisation of this result to the $\langle\cO_2\cO_2\cO_p\cO_p\rangle$ family of correlators was given in~\cite{Alday:2019nin}, where an analogous absence of single poles in the Mellin amplitude was observed.
\subsubsection*{One-loop string corrections}
The general structure of all one-loop string Mellin amplitudes $\me^{(2,k)}$ was proposed to be of the form
\begin{align}\label{eq:one-loop_Mellin_ansatz}
	\me^{(2,k)}(s,t) = \sum_{m+n=k}f^{m|n}(s,t)~\widetilde{\psi}_0\Big(2-\frac{s}{2}\Big) + f^{m|n}(t,s)~\widetilde{\psi}_0\Big(2-\frac{t}{2}\Big) + f^{m|n}(\tilde{u},t)~\widetilde{\psi}_0\Big(2-\frac{\tilde{u}}{2}\Big),
\end{align}
with the constraint
\begin{align}
	f^{m|n}(s,t) = f^{m|n}(s,\tilde{u}),
\end{align}
to ensure crossing symmetry of the full Mellin amplitude $\me^{(2,k)}(s,t)$. Instead of using the usual digamma function $\psi_0(w)$ as in~\cite{Alday:2018kkw}, we define a shifted digamma function $\widetilde{\psi}_0(w) \equiv \psi_0(w)+\gamma_E$, such that the unphysical Euler-Mascheroni constant $\gamma_E$ does not appear in the position space representation after performing the Mellin integration of $\me^{(2,k)}(s,t)$. Note that for integer values $n\in\mathbb{N}$, $\widetilde{\psi}_0(n)$ is then simply related to the harmonic numbers by $\widetilde{\psi}_0(n)=H_{n-1}$.

In the above formula~\eqref{eq:one-loop_Mellin_ansatz}, $f^{m|n}(s,t)$ is a polynomial in $s$ and $t$. The order in $s$ of this polynomial is bounded by $m+n+1$, while the order in $t$ is determined by the maximal spin contribution $\ell_{\text{max}}$ of the corresponding double discontinuity $\mathcal{D}^{m|n}$, given by\footnote{We checked that our discussion on the orders of the polynomials $f^{m|n}(s,t)$ is in agreement with the ``basis of polynomial Mellin amplitudes'' described in~\cite{Alday:2018kkw}.}
\begin{align}
\begin{split}
	\mathcal{D}^{0|n}:\quad &\ell_{\text{max}}=2\left\lfloor\frac{n-3}{2}\right\rfloor,\\ \mathcal{D}^{m|n}:\quad &\ell_{\text{max}}=2\left\lfloor\frac{\min(m,n)-3}{2}\right\rfloor.
\end{split}
\end{align}
By matching the double discontinuities $\mathcal{D}^{(0|3)}$ and $\mathcal{D}^{(3|3)}$ at orders $\ll$ and $\lambda^{-3}$, respectively, one can determine the corresponding polynomials $f^{m|n}(s,t)$ to take the forms~\cite{Alday:2018kkw}
\begin{align}
\begin{split}
	f^{0|3}(s)&= -16\zeta_3\big(63s^4-644s^3+2772s^2-5776s+4800\big),\\
	f^{3|3}(s)&= -\frac{1080\zeta_3^2}{7}\big(462s^7-11627s^6+134274s^5-908180s^4\\
			  &\qquad\qquad\quad~+3841208s^3-10071488s^2+15053056s-9838080\big),
\end{split}
\end{align}
where we made the overall normalisations consistent with our conventions. Note that both of the above amplitudes do not depend on $t$, in agreement with the spin truncation of the $\mathcal{R}^4$ vertex to spin $\ell_{\text{max}}=0$. By explicitly performing the Mellin integration in a series expansion around small $(u,v)$, we verified that the above Mellin amplitudes are in agreement with our position space results obtained by the bootstrap approach described in Section~\ref{sec:bootstrap_problem}, thus confirming the appearance of the weight-three function $f^{(3)}(\x,\xb)$.

By using the order $\lll$ data of the $\langle\cO_2\cO_2\cO_p\cO_p\rangle$ family of correlators from~\cite{Alday:2018pdi,Binder:2019jwn}, we can furthermore provide some new results. For example, we can compute the double discontinuity $\mathcal{D}^{0|5}$, resulting in
\begin{align}
\begin{split}
	f^{0|5}(s,t)&=-2\zeta_5\big(10890 s^6+45 s^5 (11 t-4669)+9 s^4 (55 t^2-640 t+204358)\\
				&\qquad\qquad-4 s^3 (945 t^2-7173 t+2285717)+36 s^2 (377 t^2-2208 t+745066)\\
				&\qquad\qquad-16 s (1575 t^2-7488 t+2722522)+576 (33 t^2-132 t+52682)\big),
\end{split}
\end{align}
which in fact appears before the $f^{3|3}(s)$ contribution in the $1/\lambda$ expansion and is the first case with non-trivial $t$-dependence. At order $\lambda^{-4}$, we can similarly compute the contribution
\begin{align}
\begin{split}
	f^{3|5}(s,t)&= -90\zeta_3\zeta_5\big(28028 s^9-1075074 s^8+19321302 s^7-211238951 s^6\\
				&\qquad\qquad~\quad+1535536842 s^5-7645987076 s^4+25938244248 s^3\\
				&\qquad\qquad~\quad-57543276224s^2+75453134080 s-44400268800\big),
\end{split}
\end{align}
and finally the order $\lambda^{-5}$ contribution from $\mathcal{D}^{5|5}$ is given by
\begin{align}
\begin{split}
	f^{5|5}(s,t)&=-\frac{45\zeta_5^2}{22}\big(57657600 s^{11}+30030 s^{10} (16 t-104093)\\
				&\qquad\qquad\quad+12012 s^9 (40 t^2-1445 t+6689071)\\
				&\qquad\qquad\quad-572 s^8 (26985 t^2-531356 t+2242111079)\\
				&\qquad\qquad\quad+22 s^7 (11008816 t^2-151917584 t+638025985123)\\
				&\qquad\qquad\quad-77 s^6 (30823520 t^2-327881344 t+1429188184721)\\
				&\qquad\qquad\quad+14 s^5 (1125229952 t^2-9688637728 t+44851775822225)\\
				&\qquad\qquad\quad-28 s^4 (2593858960 t^2-18612610496 t+92780493961669)\\
				&\qquad\qquad\quad+56 s^3 (4118587328 t^2-25104138112 t+135924547490919)\\
				&\qquad\qquad\quad-64 s^2 (7551065200 t^2-39625690048 t+234345782828097)\\
				&\qquad\qquad\quad+256 s (2355357312 t^2-10748615808 t+69677906818663)\\
				&\qquad\qquad\quad-53760 (6319936 t^2-25279744 t+180000568369)\big).
\end{split}
\end{align}
Note that the orders of the polynomials $f^{m|n}$ given above all fit into the general pattern described earlier.

Before concluding, let us mention once more that we checked agreement between our position space results and the Mellin space amplitudes described here. Such a comparison can be easily performed in a series expansion around small $(u,v)$ by using the explicit representation~\eqref{eq:f3_small_uv_expansion} of $f^{(3)}$ which is suitable for this expansion.

\section{Conclusions}\setcounter{equation}{0}
In this paper we have addressed the problem of constructing one-loop string corrections to the four-point correlation function of the stress-tensor multiplet in $\mathcal{N}=4$ SYM. In particular, we describe a position space bootstrap algorithm which relies only on the knowledge of the corresponding double discontinuity at a given order in $1/\lambda$. While we provide explicit results for the first few orders in the $1/\lambda$ expansion, the form of our one-loop ansatz is in fact valid to all orders. Our results for the final one-loop correlators are fixed up to a finite number of ambiguities, which can be fully characterised in terms of their polynomial Mellin amplitudes.

Our work complements the Mellin space approach of~\cite{Alday:2018kkw}, where a basis for the polynomials $f^{m|n}(s,t)$ was described. Therefore, at least up to one-loop, the position and Mellin space approaches are essentially interchangeable by comparison of their small $(x,\tilde{x})$ and small $(u,v)$ expansions as explained in more detail in Appendix \ref{App-f3}. The position and Mellin space approaches each have advantages and disadvantages. For example, comparably simple structures emerge when considering one-loop amplitudes in Mellin space, whereas theirspace time equivalents turn out to be rather involved. Also, the connection to ten-dimensional physics is given very directly in Mellin space through the flat-space limit, which has been explored in many references, e.g.~\cite{Penedones:2010ue,Fitzpatrick:2011ia,Fitzpatrick:2011hu, Fitzpatrick:2011dm,Goncalves:2014ffa}. On the other hand, the ansatz of transcendental functions for the position space amplitudes makes their singularity-structure very explicit, while this is quite obscure from a Mellin space point of view. In particular any analytic continuation or kinematic expansion (e.g. the OPE) that one may wish to perform is straightforward from the spacetime point of view.

Indeed, we have found that a new weight-three function is required in our ansatz, compared to the analogous bootstrap approach for one-loop supergravity correlators~\cite{Aprile:2017bgs,Aprile:2017qoy,Aprile:2019rep}. This function involves a new type of letter (or logarithmic singularity), $\x-\xb$, which is not present in the supergravity case. The appearance of $f^{(3)}(\x,\xb)$ provides a first understanding of what type of functions will appear in loop amplitudes of string theory on AdS. It would be interesting to explore whether the set of singularities found so far, namely $\{\x,\xb,1-\x,1-\xb,\x-\xb\}$, is sufficient for the description of higher loop amplitudes, or whether new letters have to be included.

As our bootstrap approach leaves a finite number of ambiguities unfixed, we have to rely on other methods in order to determine their values. One such possible avenue is provided by supersymmetric localisation, which was used in~\cite{Chester:2019pvm} to fix the single ambiguity in the supergravity result of~\cite{Aprile:2017bgs}. It would be interesting to see whether this or any other method can fix the one-loop ambiguities which arise in the $1/\lambda$ expansion.

Finally, there are further interesting open questions which can be addressed. A logical next step is to attempt the generalisation of the results presented here to correlators with more general external charges. Following the great progress for tree-level correlators~\cite{Rastelli:2016nze,Rastelli:2017udc,Alday:2018pdi,Binder:2019jwn,Drummond:2019odu}, this has already been pursued at one-loop in the supergravity case, both in position space~\cite{Aprile:2017qoy,Aprile:2019rep} and recently also in Mellin space~\cite{Alday:2019nin}. Inspiration on how to make progress in this direction might again come from tree-level supergravity, where a hidden ten-dimensional conformal symmetry provides a generating functional for the arbitrary-charge correlator in terms of the $\fourtwo$ correlator~\cite{Caron-Huot:2018kta}. In fact, it is possible to write the double discontinuities $\mathcal{D}^{m|n}$ with an eight-order differential operator $\Delta^{(8)}$ pulled out, which is reminiscent of the ten-dimensional symmetry. In the case of one-loop supergravity, this statement can be uplifted to the full correlator: up to tree-level-like terms, the one-loop correlator can be written as $\Delta^{(8)}$ acting on a much simpler ``pre-amplitude''~\cite{Aprile:2019rep}. In the recent Mellin space result of~\cite{Alday:2019nin}, this statement manifests itself in the fact that additional single poles in the Mellin amplitude are absent, such that the full Mellin amplitude is essentially determined by the double discontinuity only.\footnote{In the language of~\cite{Aprile:2019rep}, the Mellin space ansatz in terms of simultaneous poles seems to correctly reproduce the CFT data in the ``window'', without the need of additional single poles.} It would be interesting to investigate whether this differential operator can be used to generalise the construction presented here to more general correlators.

\section*{Acknowledgements}
We thank Paul Heslop for useful discussions. JMD and HP acknowledge support from ERC Consolidator grant 648630 IQFT.

\appendix
\section{Analytic properties of $f^{(3)}(x,\bar{x})$}\label{App-f3}
Here we give some more details on the structure of the function $f^{(3)}$ which makes an appearance in the one-loop string amplitudes. We recall that the total derivative is defined in equation (\ref{f3totald}). By successively stripping off the leading $\log u$ discontinuity we arrive at the form (\ref{f3loguform}) with $\tilde{f}^{(1)}$ obtained very simply and with the total derivative of $\tilde{f}^{(2)}$ obtained in the form
\begin{align}
d \tilde{f}^{(2)}(x,\bar{x}) = - &\bigl[2 \log(1-x) \bigr] d \log x  \notag \\
+ &\bigl[ 2 \log(1-\bar{x}) \bigr] d\log \bar{x}  \notag \\
- & \bigl[  \log(1-x) - 3 \log (1-\bar{x}) \bigr] d \log (1-x) \notag \\
+ & \bigl[\log(1-\bar{x})  -  3 \log (1-x) \bigr] d \log (1-\bar{x}) \notag \\
+ & \bigl[6 \log(1-x) - 6 \log(1-\bar{x}) \bigr] d \log (x-\bar{x}) \,.
\end{align}
The form (\ref{f2texp}) agrees with the above and obeys $\tilde{f}^{(2)}(x,x)=0$ as it should by antisymmetry.
Finally we obtain the total derivative of $\tilde{f}^{(3)}$ in the form
\begin{align}\label{dft3}
d \tilde{f}^{(3)}(x,\bar{x}) = \quad &\bigl[-4({\rm Li}_2(x) - {\rm Li}_2(\bar{x})) +\tfrac{1}{2}\log^2 v  - \tilde{f}^{(2)}(x,\bar{x})  \bigr] d \log x  \notag \\
+ &\bigl[-4({\rm Li}_2(x) - {\rm Li}_2(\bar{x})) -\tfrac{1}{2}\log^2 v - \tilde{f}^{(2)}(x,\bar{x}) \bigr] d\log \bar{x}  \notag \\
+ & \bigl[-4({\rm Li}_2(x) - {\rm Li}_2(\bar{x})) \bigr] d \log (1-x) \notag \\
+ & \bigl[-4({\rm Li}_2(x) - {\rm Li}_2(\bar{x})) \bigr] d \log (1-\bar{x}) \notag \\
+ & \bigl[12({\rm Li}_2(x) - {\rm Li}_2(\bar{x})) \bigr] d \log (x-\bar{x}) \,.
\end{align}
We can easily integrate this in a form suitable for expansion in small $x$ and $\bar{x}$. However for comparison to Mellin space it is more convenient to make the change of variables $\tilde{x} = 1- \bar{x}$ so that
\be
u=x(1-\tilde{x})\,, \qquad v=\tilde{x}(1-x)\,.
\ee
Then the small $x$ and $\tilde{x}$ expansion can easily be compared to a small $u$ and $v$ expansion.

To this end we first pull (\ref{dft3}) back to the line $x=0$,
\be
d \tilde{f}^{(3)}(0,\bar{x}) = \bigl[-12 {\rm Li}_2(\bar{x}) -\log^2 (1-\bar{x}) \bigl] d \log \bar{x} + \bigl[4 \,{\rm Li}_2(\bar{x})\bigr] d \log (1-\bar{x})\,.
\ee
This can be easily integrated in terms of weight three harmonic polylogarithms \cite{Remiddi:1999ew} with the condition that $\tilde{f}^{(3)}(0,0)=0$,
\be
\tilde{f}^{(3)}(0,\bar{x}) = -12 H_{3}(\bar{x}) - 4 H_{1, 2}(\bar{x}) - 2 H_{2, 1}(\bar{x})\,.
\ee
Now performing our change of variables from $\bar{x}$ to $\tilde{x}=1-\bar{x}$ we have in the small $\tilde{x}$ expansion,
\be
\tilde{f}^{(3)}(0,1-\tilde{x}) = -6\zeta_3 + 4 \zeta_2 \log \tilde{x} + O(\tilde{x})\,.
\label{zeta3fix}
\ee
Now using
\begin{align}
 \phi^{(1)}(x,1-\tilde{x}) = &- \log u \log v -2\bigl[ {\rm Li}_1(x) \log u + {\rm Li}_1(\tilde{x}) \log v\bigr] \notag \\
 &-2\bigl[\zeta_2 + {\rm Li}_1(x) {\rm Li}_1(\tilde{x}) - {\rm Li}_2(x) - {\rm Li}_2(\tilde{x})\bigr]  \,,
\end{align}
we may write 
\begin{align}
df^{(3)}(x,1-\tilde{x}) = \quad &\bigl[- 2 \phi^{(1)}(x,1-\tilde{x}) +\tfrac{1}{2}\log^2 v - \log u \log v  \bigr] d \log x  \notag \\
+ &\bigl[- 2 \phi^{(1)}(x,1-\tilde{x}) -\tfrac{1}{2}\log^2 v + \log u \log v \bigr] d\log (1-\tilde{x})  \notag \\
+ & \bigl[- 2 \phi^{(1)}(x,1-\tilde{x}) -\tfrac{1}{2}\log^2 u + \log u \log v \bigr] d \log (1-x) \notag \\
+ & \bigl[- 2 \phi^{(1)}(x,1-\tilde{x}) +\tfrac{1}{2}\log^2 u - \log u \log v \bigr] d \log \tilde{x}  \notag \\
+ & \bigl[6\phi^{(1)}(x,1-\tilde{x})\bigr] d \log (1-x-\tilde{x}) \,.
\end{align}
We can then make manifest all the logarithmic singularities in $\log u$ and $\log v$ as follows,
\begin{align}\label{eq:f3_small_uv_expansion}
f^{(3)}(x,1-\tilde{x}) =  &\quad \frac{1}{2} \log^2 u \log v + \frac{1}{2} \log u \log^2 v -\log^2 u \log(1-x) - \log^2v \log(1-\tilde{x}) \notag \\ &
+\log u \log v~ \bigl[2 \log(1-x) + 2 \log(1-\tilde{x}) - 6 \log (1-x-\tilde{x})\bigr] \notag \\
&+ \log u ~g^{(2)}(x,\tilde{x}) + \log v ~g^{(2)}(\tilde{x},x) + 6g^{(3)}(x,\tilde{x})\,,
\end{align}
where the function $g^{(2)}$ can be expressed as
\be
g^{(2)}(x,\tilde{x}) = 4 \zeta_2 +2 \,{\rm Li}_{2}(x) + 2\, {\rm Li}_{2}(\tilde{x}) - 6\, {\rm Li}_{2}\Bigl(\frac{\tilde{x}}{1 - x}\Bigr) - 2\log(1-x) \log \frac{(1-x)^3 (1-\tilde{x})}{(1-x-\tilde{x})^6}\,.
\ee
To write a formula for $g^{(3)}$ it is helpful to use hyperlogarithms, $G_w(t)$ which depend on a word $w=a_1 a_2 \ldots a_n$ in letters $a_i$ and a variable $t$. The function whose word is just a string of $n$ zeros is a power of $\log t$,
\be
G_{0^n}(t) = \frac{1}{n!} \log^n t\,.
\ee
The other functions are defined recursively,
\be
G_{aw}(t) = \int_0^t \frac{ds}{s-a} G_w (s)\,.
\ee
Using these hyperlogarithms we can write an expression for $g^{(3)}$ by integrating the total derivative and fixing the term proportional to $\zeta_3$ from (\ref{zeta3fix}),
\begin{align}
g^{(3)}(x,\tilde{x}) = & \quad\,\, G_{1}(\tilde{x}) G_{0, 1}(x) - 2 G_{1 - x}( \tilde{x}) G_{0,1}(x) -  2 G_{1}(\tilde{x}) G_{1, 1}(x) + G_{1}(x) G_{1, 1}(\tilde{x}) \notag \\
&- 2 G_{1}(x) G_{1, 1 - x}(\tilde{x}) 
 - 2 G_{1}(x) G_{1 - x, 1}(\tilde{x}) + 
 G_{0, 0, 1}(x) + G_{0, 0, 1}(\tilde{x}) 
 \notag \\
 &- 2G_{0, 1, 1}(x) + 
 G_{0, 1, 1}(\tilde{x}) - G_{0, 1, 1 - x}(\tilde{x}) - 
 2 G_{0, 1 - x, 1}(\tilde{x}) - 2 G_{1, 0, 1}(x) \notag \\
 &+ G_{1, 0, 1}(\tilde{x}) - 
 G_{1, 0, 1 - x}(\tilde{x}) - 2 G_{1 - x, 0, 1}(\tilde{x}) -2\zeta_2 \log(1-x-\tilde{x}) - \zeta_3 \,.
\end{align}
Although it is not manifest from the above formula $g^{(3)}$ is symmetric, $g^{(3)}(x,\tilde{x}) = g^{(3)}(\tilde{x},x)$. The apparent asymmetry is simply due to a choice of the contour of integration (first in the $x$ direction, then the $\tilde{x}$ direction).


\end{document}